\documentclass[journal, twocolumn, 10pt]{IEEEtran}

\usepackage{amssymb,latexsym,amsfonts,amsmath,amsthm}
\usepackage{graphicx}
\usepackage{xcolor}
\usepackage{hyperref}	
\usepackage{cleveref}
\usepackage{subcaption}

\usepackage{algorithm}
\usepackage{algorithmicx}
\usepackage[noend]{algpseudocode}

\usepackage{tikz}
\usetikzlibrary{arrows, patterns, math, positioning}
\usepackage{pgfplots}

\usepackage{blkarray}
\usepackage{tabularx}
\usepackage{multirow}
\usepackage{enumitem}

\usepackage{cite}
\ifCLASSINFOpdf
\else
\fi
\usepackage{amsmath}
\usepackage{url}
\newcommand\todo[1]{\textcolor{red}{\textbf{#1}}}

\newcommand{\XX}{x}%{{\mathbb{X}}}
\newcommand{\UU}{u}%{{\mathbb{U}}}
\newcommand{\WW}{w}%{{\mathbb{W}}}
\newcommand{\YY}{y}%{{\mathbb{Y}}}
\newcommand{\ZZ}{z}%{{\mathbb{Z}}}
\newcommand{\OO}{o}%{\mathbb{O}}
\newcommand{\II}{i}%{\mathbb{I}}
\newcommand{\W}{w}%{\mathbb{W}}
\newcommand{\V}{v}%{\mathbb{V}}

\newcommand{\sys}{{F}}
\newcommand{\maps}{\rightarrow}
\newcommand\quant[1]{Q_{#1}}
\newcommand\abs[1]{\hat{#1}}
\newcommand\NB[1]{{NB}_{#1}}%{\textsf{NB}_{#1}}

\newcommand{\true}{\top}
\newcommand{\false}{\bot}
\newcommand{\NOT}{\neg}
\newcommand{\OR}{\vee}
\newcommand{\AND}{\wedge}
\renewcommand{\implies}{\Rightarrow}
\newcommand{\vimplies}{\Downarrow}
\newcommand\equant[1]{\exists #1}
\newcommand\uquant[1]{\forall #1}

\newcommand\dom[1]{\mathcal{D}(#1)}
\newcommand{\sequiv}{\equiv}
\newcommand{\vequiv}{\sequiv}
\newcommand{\lequiv}{\Leftrightarrow}

\newtheorem{definition}{Definition}
\newtheorem{example}{Example}
\newtheorem{theorem}{Theorem}

\newcommand{\OA}{OA}
\newcommand{\modabs}{\preceq}
\newcommand{\comp}{\textsf{Comp}}

\newcommand\pow[1]{2^{#1}}
\newcommand\pref[1]{(\ref{#1})}

\renewcommand{\emptyset}{{\varnothing}}

\begin{document}
\title{Abstractions for Symbolic Controller Synthesis are Composable} 

%
% author names and IEEE memberships
% note positions of commas and nonbreaking spaces ( ~ ) LaTeX will not break
% a structure at a ~ so this keeps an author's name from being broken across
% two lines.
% use \thanks{} to gain access to the first footnote area
% a separate \thanks must be used for each paragraph as LaTeX2e's \thanks
% was not built to handle multiple paragraphs
%

\author{Eric~S.~Kim, 
        Murat~Arcak

\thanks{The authors are with the Department of Electrical Engineering and Computer Sciences at the University of California, Berkeley, CA, USA. 
}% 
}% <-this % stops a space
%\thanks{J. Doe and J. Doe are with Anonymous University.}% <-this % stops a space
%\thanks{Manuscript received April 19, 2005; revised August 26, 2015.}

% note the % following the last \IEEEmembership and also \thanks - 
% these prevent an unwanted space from occurring between the last author name
% and the end of the author line. i.e., if you had this:
% 
% \author{....lastname \thanks{...} \thanks{...} }
%                     ^------------^------------^----Do not want these spaces!
%
% a space would be appended to the last name and could cause every name on that
% line to be shifted left slightly. This is one of those "LaTeX things". For
% instance, "\textbf{A} \textbf{B}" will typeset as "A B" not "AB". To get
% "AB" then you have to do: "\textbf{A}\textbf{B}"
% \thanks is no different in this regard, so shield the last } of each \thanks
% that ends a line with a % and do not let a space in before the next \thanks.
% Spaces after \IEEEmembership other than the last one are OK (and needed) as
% you are supposed to have spaces between the names. For what it is worth,
% this is a minor point as most people would not even notice if the said evil
% space somehow managed to creep in.

% The paper headers
\markboth{In Submission}%
{\todo{Title here}}
% The only time the second header will appear is for the odd numbered pages
% after the title page when using the twoside option.
% 
% *** Note that you probably will NOT want to include the author's ***
% *** name in the headers of peer review papers.                   ***
% You can use \ifCLASSOPTIONpeerreview for conditional compilation here if
% you desire.

% make the title area
\maketitle

% As a general rule, do not put math, special symbols or citations
% in the abstract or keywords.
\begin{abstract} 
Translating continuous control system models into finite automata allows us to use powerful discrete tools to synthesize controllers for complex specifications. The abstraction construction step is unfortunately hamstrung by high runtime and memory requirements for high dimensional systems. 
This paper describes how the transition relation encoding the abstract system dynamics can be generated by connecting smaller abstract modules in series and parallel. We provide a composition operation and show that composing a collection of abstract modules yields another abstraction satisfying a feedback refinement relation. Through compositionality we circumvent the acute computational cost of directly abstracting a high dimensional system and also modularize the abstraction construction pipeline. 
\end{abstract}

% Note that keywords are not normally used for peerreview papers.

% For peer review papers, you can put extra information on the cover
% page as needed:
% \ifCLASSOPTIONpeerreview
% \begin{center} \bfseries EDICS Category: 3-BBND \end{center}
% \fi
%
% For peerreview papers, this IEEEtran command inserts a page break and
% creates the second title. It will be ignored for other modes.
%\IEEEpeerreviewmaketitle

%\input{todo.tex}
\section{Introduction}

\IEEEPARstart{T}{he} high level goal of controller synthesis is the automatic construction of control software that causes a closed loop system to satisfy a desirable specification such as safety, reachability, or recurrence. 
%For systems with complex dynamics and nonconvex constraint set, it is not always feasible to leverage compact algebraic descriptions that are available for linear and polynomial dynamical systems. 
Gridding the state space and explicitly reasoning about sets of behaviors is a general approach that can accommodate nonlinear dynamics and specifications over complex sets with temporal dependencies.
Recent literature has extensively investigated methods to systematically construct finite abstractions, which are automata whose states arise from a partition of the underlying continuous state space and whose transitions are consistent with the continuous dynamics \cite{Tabu, majid}.
One commonly cited reason holding back wider adoption of this approach is that the abstraction algorithm quickly runs into
computational bottlenecks with respect to the system state-input dimensions.
%Our prior work \cite{sparseabs2017} \cite{Kim:2018:CCS:3178126.3178144} has shown that dimensionality alone is an inadequate metric for estimating the computational complexity of abstraction and synthesis. Leveraging information about the internal interconnection topology among components can lead to a dramatic reduction in the runtime and memory requirements for constructing abstractions. 

%\subsection{Goal: Modularizing the Finite Abstraction Pipeline}
%\begin{figure}
%\centering
%\includegraphics[width = \columnwidth]{figures/modlibrary.png}\\
%\includegraphics[width = \columnwidth]{figures/composedmods.png}\\
%\caption{\small From a library of concrete modules and their associated finite abstractions, we construct an abstraction of the concrete system $x' :=  \log(K|x|+1)+u$ by composing abstract components. The linker handles issues like variable renaming. The new variables flowing along each wire are discretized variables instead of continuous ones, e.g. $h$ changes to $\abs h$.}
%\end{figure}

We advocate for a bottom-up approach by composing and interconnecting small modules to encode the control system dynamics. Graphical modeling languages such as Simulink make use of a similar principle where complex system behaviors emerge via the interaction of simple components. We break the control system abstraction procedure into two steps: 
\begin{enumerate}[leftmargin=4.5mm]
\item \textbf{Abstract Individual Modules:} Generate a discrete abstraction for each individual module in a library. 
\item \textbf{Compose Modules:} Compose modules to create a larger module that encodes the control system dynamics.  
\end{enumerate}
Benefits of modularizing the abstraction pipeline include: 
\begin{itemize}[leftmargin=3.5mm]
\item \textbf{Efficiency:} Abstracting lower dimensional components reduces computational requirements. 
\item \textbf{Reuse and Portability:} Abstract modules can be reused in different contexts by interconnecting them after abstraction. 
\item \textbf{Compartmentalization:} Modifying or swapping a module does not require re-abstracting other modules. 
\item \textbf{Specialization:}  Properties such as mixed-monotonicity \cite{coogan} make it easier to construct tight system abstractions which are favorable for controller synthesis.
These properties may be present on the level of individual modules, but may not hold globally across all modules in a system.
\end{itemize}

\subsection{Contributions}

This paper's main contribution is a theorem that permits abstract components to be composed while preserving a feedback refinement relationship between abstract and concrete systems. 
Feedback refinement relations \cite{gunther2} certify that an abstraction mimics the dynamics of the concrete control system and account for the information loss incurred from discretizing the state and input spaces. They ensure that a controller designed to enforce some behavior over the abstraction can be translated into a controller enforcing the same behavior over the original continuous system, modulo an approximation parameter induced by the state space grid.
%We show that the feedback refinement relation is a specialized version of a similar but more general relation which we call an approximate module abstraction. 
The composition operation only imposes mild requirements on the interconnection topology alone; it forbids circular dependencies without a delay (i.e., an algebraic loop) and connections between modules must satisfy a simple type checking constraint. The composition theorem generalizes prior work from \cite{sparseabs2017} \cite{Kim:2018:CCS:3178126.3178144} that allowed limited forms of interconnections and can augment existing tools for finite abstraction and controller synthesis such as \texttt{SCOTS}\cite{Rungger2016}.

\Cref{sec:notation} motivates the use of logical descriptions of control systems instead of as functions or set valued maps. 
\Cref{s:modules} formally defines modules and a relationship that encodes the connection between concrete and abstract modules. Feedback refinement relations are shown to be a special case of this relationship. \Cref{s:absconstruction} describes how finite abstractions are represented and constructed algorithmically.
\Cref{s:modcomposition} defines a generic module composition operator and contains our main technical contribution showing that the concrete-abstract system relationship is preserved under this operator. We contrast our work with literature focusing on decentralized symbolic controller synthesis in \Cref{sec:related}.

\subsection{Illustration of System Decomposition into Modules} \label{ss:sysdecomp}
%\begin{figure}
%\begin{algorithmic}[1] % The number tells where the line numbering should start
%\Function{$\mathcal{F}$}{$\code{x\_1, x\_2, u\_1, u\_2}$}
%\State $\code{float l\_1 := 3*x\_1}$
%\State $\code{float l\_2 := sqrt(l\_1)}$
%\State $\code{float l\_3 := x\_2 * l\_2}$
%\State $\code{float l\_4 := u\_2 * u\_2}$
%\State $\code{float x\_1' := l\_2 - u\_2 + u\_1}$
%\State $\code{float x\_2' := l\_3 + l\_4}$
%\State \Return $\code{x\_1', x\_2'}$
%\EndFunction
%\end{algorithmic}
%\caption{\small Psuedocode for the control system in \Cref{eq:simulink_eqn}.} \label{alg:simulinkpsuedocode}
%\end{figure}

\begin{figure}
\centering
\includegraphics[width=.9\columnwidth]{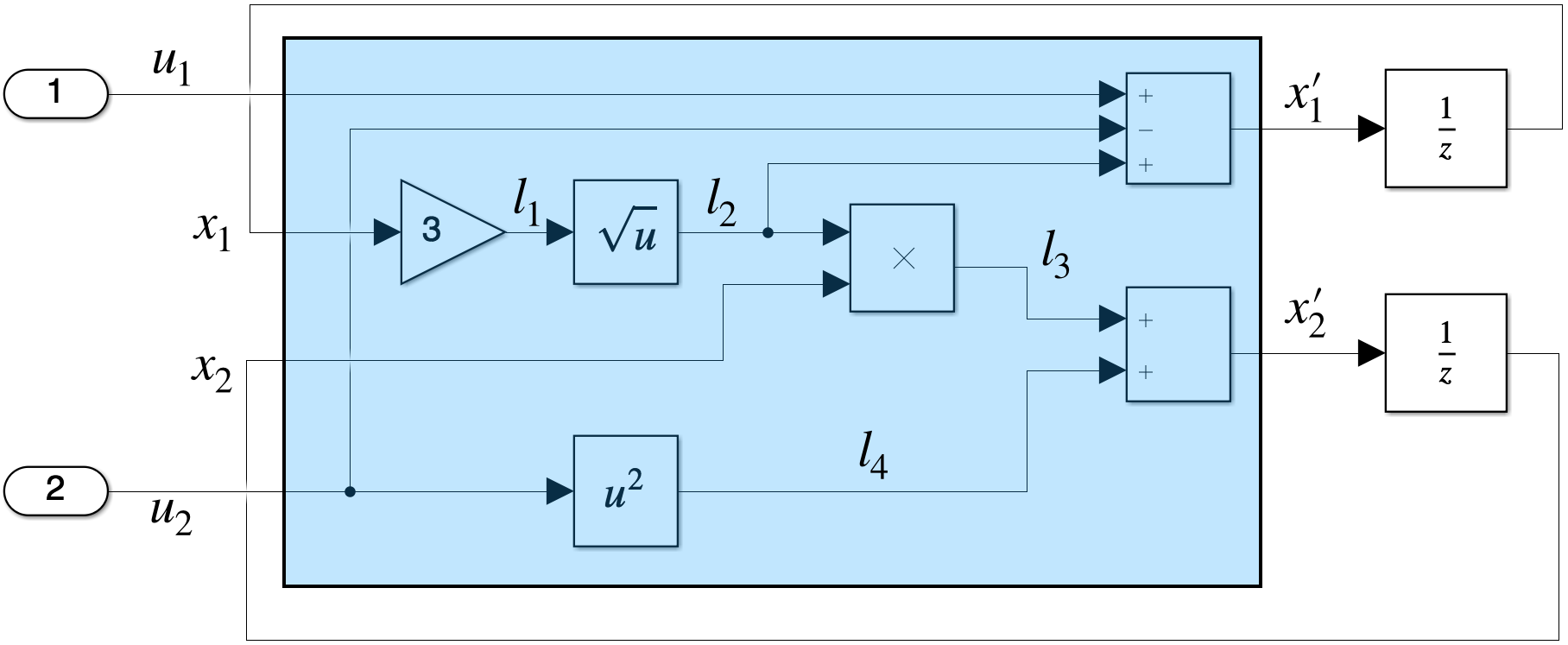}
\caption{\small Simulink visualization of \Cref{eq:simulink_eqn}. The blue box is a composite module consisting of internal modules. } \label{fig:simulinkex}
\end{figure}

Consider as an example the discrete time control system 
\begin{align}
x' := \mathcal{F} (x,u) \label{eqn:monolithicsys}
\end{align}
with two continuous current states $x_1, x_2 \in \mathbb{R}$, two control inputs $u_1, u_2 \in \mathbb{R}$, and two next states $x_1',x_2' \in \mathbb{R}$. Let $\mathcal{F}(x,u)$ be concretely given by \Cref{eq:simulink_eqn}. 
\begin{align}
\left(
\begin{array}{l}
x_1' \\
x_2'
\end{array}\right)
:=
\left(\begin{array}{l}
\sqrt{3 x_1} - u_2 + u_1\\
 x_2 \sqrt{3 x_1} +  u_2^2
\end{array}\right)\label{eq:simulink_eqn} 
\end{align} 
The encodings (\ref{eqn:monolithicsys}) directly maps the variables $u_1, u_2, x_1, x_2$  to a next state $x_1', x_2'$, but this obfuscates some internal structure that is more clear when the system is viewed as the blue box in \Cref{fig:simulinkex}.
Latent variables $l_1 := 3x_1$,  $l_2 := \sqrt{3x_1}$, $ l_3 :=x_2 \sqrt{3x_1}$, and $l_4 := u_2^2$ capture intermediate computations, with $l_2$ is also shared across updates for both $x_1'$ and $x_2'$.  
%These latent variables correspond with wires that are fully contained inside the blue box or the function body's internal variables.

Modules are analogous to the blue box in \Cref{fig:simulinkex} and the blocks it contains. They are memoryless and encode the system's transition relation.
Two modules are connected in series if the output of one module feeds into the input of another, such as the gain and square root modules. Two components are connected in parallel if neither output is connected to the other module's input, such as the summation components. Module composition yields another module. 
In our framework, unit delay blocks are not considered modules. They only appear to break any algebraic loops and introduce state in the context of controller synthesis, but they are not required for encoding the dynamics $\mathcal{F}(x,u)$. 
%While modules are memoryless, they can still be used to describe system dynamics.
%The control system in \Cref{eq:simulink_eqn} contains variables $x_1, x_2, x_1', x_2'$ that are associated with states, 
%\Cref{fig:simulinkex}'s blue box does not contain any internal delay blocks; 

\section{Logical System Representations and Notation}\label{sec:notation}

%In graphical programming languages such as Simulink, variables are wires which carry a time varying signal and modules are connected by sharing a wire. Modules are components that impose coupling constraints across multiple variables.
%Each wire can also carry multiple signals at once.

%\subsection{Representing Discrete Time Control Systems with Predicates} 

The assignment operator ``:=" in \Cref{eq:simulink_eqn} is problematic because the example can exhibit non-determinism and blocking behaviors, which commonly appear in finite state abstractions as will be shown in \Cref{s:absconstruction}.
Under the system dynamics, $x_1$ can easily become negative but this induces the system to block because $x_1'$ and $x_2'$ are calculated by taking the square root $\sqrt{3x_1}$ (we ignore complex numbers). 
Even if $x_1 \geq 0$, the square root module may be non-deterministic and output either a positive or negative square root. 
%In this case, \Cref{eq:simulink_eqn} should be rewritten as the set membership constraint
%\begin{align}
%\left( 
%\begin{array}{l}
%x_1' \\
%x_2'
%\end{array}\right)
%\in 
%\left\{ \begin{array}{r}
% l_2 - u_2 + u_1  \\
% x_2 l_2 +  u_2^2 
%\end{array} \Big| l_2 \in \left\{ \begin{array}{c}
%|\sqrt{3 x_1}|,\\ - |\sqrt{3 x_1}| 
%\end{array}\right\}  \right\}. \label{eq:simulinkinclusion}
%\end{align} 

Predicates can accommodate both non-determinism and undefined outputs in a unified notation. Predicates are functions that output a Boolean value and can be interpreted as set indicator functions or as constraints to be satisfied.
We can replace $x' := \mathcal{F}(x,u)$ with a predicate representation $F(x,u,x')$, which only accepts those values of $x,u,x'$ where for some $l_2$ satisfy each of the following:
\begin{align*}
x_1' &== l_2 - u_2 + u_1 \\
x_2' &==  x_2 l_2 +  u_2^2 \\
l_2 &\in \{ |\sqrt{3x_1}|, -|\sqrt{3x_1}|\} \\
x_1 &\geq 0.
\end{align*}
\Cref{s:modcomposition} on module composition will formally show how constraints like those above are generated. 
%Replacing the assignment operator ``$:=$" in \Cref{eq:simulink_eqn} with an equality operator ``$==$" yields its corresponding predicate.
%\begin{align*} 
%(x_1' == \sqrt{3 x_1} - u_2 + u_1) \AND 
%(x_2' ==  x_2 \sqrt{3 x_1} +  u_2^2)
%%\label{eq:simulink_predicate}
%\end{align*} 

%Replacing each ``$:=$" with ``$==$" in lines 2-7 in \Cref{alg:simulinkpsuedocode} yields a collection of six simpler predicates.
%\begin{align*}
%(x_1' == l_2 - u_2 + u_1) \AND 
%(x_2' ==  x_2 l_2 +  u_2^2) \AND
%(l_2 == \sqrt{3x_1})
%\end{align*}
%The expression (\ref{eq:simulinkinclusion}) for set-valued dynamics is a predicate expressed as a set inclusion constraint. 

\subsection{Manipulating Predicates}
{
We briefly introduce predicates and the operations used to manipulate them. A formal introduction is provided in \cite{huth2004logic}. 

Let $\true$ denote logical true and $\false$ denote false. Operators $\NOT, \AND, \OR$ respectively represent negation, conjunction, and disjunction. The implication $ a \implies b $ is a shorthand for the formula $\NOT a \OR b$. %, and $a \iff b$ is identical to $(a \implies b) \AND (b \implies a)$.
We use the assignment notation $:=$ for new definitions or declarations. The $==$ operator is a generic equivalence check between two objects of the same type and returns either true or false. Special cases of the equivalence check are set equivalence $\sequiv$ and logical equivalence $\lequiv$. 

Variables are denoted by lower case letters and are analogous to wires in Simulink. Each variable $v$ is associated with a domain of values $\dom{v}$, which is analogous to the variable's type. A composite variable is a set of variables and is analogous to a bundle of wires held together by tie wraps. The composite variable $\V$ can be constructed by taking a union $\V := \V_1 \cup \ldots \cup \V_M$ and the domain $\dom{\V} \equiv \prod_{i=1}^M \dom{\V_i}$.  For example if $v$ is associated with a $M$-dimensional Euclidean space $\mathbb{R}^M$, then it is a composite variable that can be broken apart into a collection of atomic variables $v_1, \ldots, v_M$ where $\dom{v_i} \sequiv \mathbb{R}$ for all $i := 1,\ldots,M$. All technical results herein do not distinguish between composite and atomic variables. 

Predicates are functions that map variable assignments to a Boolean value. Boolean valued expressions like ``$x \in \{4,5,12\}$" and ``$y == \sin(x)$" are predicates. The variables contained in those expressions are unassigned in the sense that they are not associated with a single value. Once all of a predicate's variables are assigned, it returns a Boolean value. Predicates without full  variable assignments yield newer predicates, e.g. assigning $y=1$ in ``$(y == \sin(x))$" yields the predicate ``$(1 == \sin(x))$". Assignment of a composite variable $v \sequiv v_1 \cup \ldots \cup v_M$  means that every $v_i$ is assigned to an element in $\dom{v_i}$.
Predicates that stand in for expressions are denoted by capital letters and are often written with the variables that appear within them, e.g. a predicate $P(v,w)$ can stand in for the expression $``v \leq w"$. 

Predicate variables are omitted when necessary in order to avoid bloated expressions and notation overhead. For example, $P(v,w)$ can simply be denoted by $P$ when clear from context that it is associated with $v$ and $w$.
Moreover, if some variables are composite then the notation can be expanded. That is, if $v \vequiv v_1 \cup \ldots \cup v_M$ then $P(v_1, \ldots, v_M, w)$ and $P(v,w)$ represent the same predicate.

Predicates can construct sets via set builder notation. A single predicate can instantiate different sets if the domains differ, e.g. $\{ x \in \dom{\XX} | P(x)\}$ and $\{(x,y) \in \dom{\XX} \times \dom{\YY} | P(x)\}$ are distinct sets but are associated with the same predicate.

The standard Boolean operations can be applied to a predicate's Boolean output to construct new predicates. The negated predicate $\neg P(v)$ is true for an assignment to $v$ if and only if $P(v)$ is false. The domain of a predicate obtained via a binary operation is the union of the two variable domains, e.g., conjunction $P(v) \AND Q(w)$ yields a predicate $(P \AND Q)(v,w)$.
Predicates $P$ and $Q$ are logically equivalent (denoted by $P \lequiv Q$) if and only if $P \implies Q$ and $Q \implies P$ are true for all variable assignments.

%For instance, the predicates $P(x)$ and $Q(x,y)$ below are logically equivalent when $\YY$ is a Boolean variable \begin{align*} 
%Q(x,y) \lequiv \big( P(x) \AND (y \OR \neg y) \big) \lequiv \big( P(x) \AND \true \big) \lequiv  P(x). 
%\end{align*} 

The universal quantifier $\forall$ and existential quantifier $\exists$ eliminate predicates variables and are analogous to set projection operations.
Given predicate $P(v,w)$, $\equant{\W}P$ and $\uquant{\W}P$ are predicates over $v$. An assignment $v \in \dom{v}$ satisfies $\equant{\W}P$ if and only if there exists an assignment $w \in \dom{w}$ such that $P(v,w)$ evaluates to true. 
Similarly, an assignment to $v$ satisfies $\uquant{\W}P$   if and only if $P(v,w)$ evaluates to true for all assignments $w \in \dom{w} $.
Applying DeMorgan's law yields the identities $\neg \equant{\W}P \lequiv \uquant{\W}(\neg P)$ and $\neg \uquant{\W}P \lequiv \equant{\W}(\neg P)$.
If the variable to be eliminated does not exist in the predicate, then the same predicate is returned. 
If the variable $\W$ is actually a composite variable $\W := \W_1 \cup \ldots \cup \W_N$ then $\equant{\W}P$ is simply a shorthand for $\equant{\W_1}\ldots\equant{\W_N} P$.

%The predicate $\forall x(Q(x) \implies P(x))$ encodes the statement ``$P(x)$ is true for every $x$ that satisfies $Q(x)$". If $Q$ and $P$ are interpreted as subset of a fixed domain, it is equivalent to the assertion $Q \subseteq P$.

}

\section{Modules and Their Abstractions} \label{s:modules}

Modules are represented as predicates where each variable is assigned a role as an input or output. 
\begin{definition}[Modules]  
A module is a triple $(\II,\OO,F)$ where $\II$ is a set of input variables, $\OO$ is a set of output variables, and predicate $F(i,o)$ is a joint constraint on variables $\II$ and $\OO$. 
\end{definition}

For a module that implements a scaling function from input $i$ to output $o$ with fixed gain $\kappa \in \mathbb{R}$, the predicate is given by an equality condition $(o == \kappa i)$. After assigning concrete values to $o$ and $i$,  this predicate evaluates to $\true$ or $\false$.

Control system modules are a special kind of module with state variables $x,x'$ satisfying $\dom{\XX} \sequiv \dom{\XX'}$ and a controllable input variable $u$.
\begin{definition}[Control System Module]
A time invariant discrete time control system is represented by the module $(\XX \cup \UU, \XX', F)$. \label{def:controlsys} 
Variables $\XX, \UU, \XX'$ can be composite so the system state and control inputs may be multi-dimensional. 
\end{definition}

An input's assignment is blocking if all possible outputs violate the module's predicate.
\begin{definition}[Nonblocking Inputs] \label{def:NB}
Module $(\II,\OO,F)'s$ nonblocking inputs are denoted by the predicate $\NB{F}(i) := \equant{\OO}F(i,o)$. The blocking inputs are $\neg \NB{F}$ or equivalently $\uquant{\OO}(\neg F)$.
\end{definition}

Consider the module $ (x,y, \; y == \sqrt{x})$. The nonblocking predicate $\equant{\YY}(y == \sqrt{x})$ is equivalent to the input constraint $x \geq 0$ because no assignment to variable $y$ can be equivalent to an undefined value when $x$ is negative.
Similarly, for the division module $(x \cup y, z, z == \frac{x}{y})$ the nonblocking predicate $\NB{z == \frac{x}{y}}$ reduces down to the expression $y \neq 0$.

%\subsection{Approximate Variable and Module Abstractions}
%\label{s:abstractions}

\subsection{Approximate Variable Abstraction}

Constructing finite abstractions starts with defining a finite domain that is related to the continuous domain. 
%Let each variable $\XX, \UU, \LL$ have an associated quantized or discrete counterpart $\abs{\XX}, \abs{\UU}, \abs{\LL}$. 
Consider a generic concrete variable $\WW$; we denote its associated abstract variable as $\hat{\WW}$.
The quantization relation $\quant{\WW}(w,\abs w)$ is a predicate that formalizes the relationship and evaluates to true whenever a pair $w,\abs w$ is related. 
This quantization predicate is called strict if for all assignments to $w$ there exists an assignment to $\abs w$ such that $\quant{\WW}(w, \abs w)$ evaluates to true. More succinctly, the predicate $\uquant{w}\equant{\abs w} \quant{\WW}(w, \abs w) \lequiv \true$.
The relation may be interpreted as a pair of set-valued maps $\quant{\WW}: \dom{w} \maps 2^{\dom{\abs \WW}}$ and $\quant{\WW}^{-1}: \dom{\abs \WW} \maps 2^{\dom{\WW}}$. From this point of view, the relation is strict if and only if $\quant{\WW}(w) \neq \emptyset$ for all assignments to $w$ or, alternatively, if $\cup_{\abs w \in \dom{\abs \WW}} \quant{\WW}^{-1}(\abs w)$ is a cover of $\dom{\WW}$.
For quantization relations over multiple variables such as $\quant{\XX \cup \YY}(x,y,\abs x,\abs y)$, we adopt a convention where the relation is decomposed component-wise 
\begin{align}
\quant{\XX \cup \YY}(x,y,\abs x,\abs y) \lequiv \quant{\XX}(x,\abs x) \AND \quant{\YY}(y,\abs y). \label{eq:decompquant}
\end{align}

A common space-discretization pair is a bounded subset of Euclidean space $\mathbb{R}^N$ paired with a finite cover of hyperrectangles. For example, the quantization relation $\quant{\WW}$ 
\begin{align}
\quant{\WW}(w, \abs w) := \left(||w - \abs w||_\infty \leq \frac{\eta}{2}\right)
\end{align}
encodes a cover that consists of infinity-norm balls of diameter $\eta > 0 $ for a domain $\dom{\abs w} \equiv \{ \eta x| x \in \mathbb{Z}^N \}$ of discrete points. 

Occasionally a concrete variable is already discrete and doesn't need to be abstracted. In this scenario $w$ and $\abs w$ will refer to the same set of variables and we use the identity relation for $\quant{\WW}(w, \abs w)$. It is trivially $\true$ because $w,\abs w$ can only be assigned to the same value. %In this scenario, the quantization relation $\quant{\WW}(w,\abs w) := (w == \abs w)$ is satisfied whenever the variables match.

\subsection{Module Approximations} 

Once each concrete variable is associated with an abstract counterpart, we can establish relationships between abstract and concrete modules. 

\begin{definition}[Approximate Module Abstraction] \label{def:relabs} 
Let $(i,o,M)$ and $(\abs i, \abs o, \abs M)$
%\begin{align*}
%M&: \II \cup \OO \maps \bool\\
%\hat{M}&: \abs {\II} \cup \abs {\OO} \maps \bool .
%\end{align*}
be modules and $\quant{\II}$ and $\quant{\OO}$ be strict quantization relations. $\hat{M}$ is an approximation of $M$ with respect to $\quant{\II}$ and $\quant{\OO}$ if and only if substituting predicates $M$ and $\abs M$ into
\begin{align}
(\quant{\II}(i,\abs i) \AND \NB{\abs M}(\abs i)) \implies \NB{M}(i) \label{eqn:NB}
\end{align}
and 
\begin{align} 
\left(\begin{array}{c}
\quant{\II}(i,\abs i) \AND \NB{\abs M}(\abs i)  \AND M(i,o) \AND \quant{\OO}(o,\abs o)\\
 \vimplies \\
  \hat{M}(\abs i, \abs o ) 
\end{array}\right)
 \label{eqn:superset}
\end{align} 
yields predicates equivalent to $\true$ for all variable assignments. This relationship is denoted by $\abs M \modabs_{\II, \OO} M$. 
\end{definition}

\Cref{def:relabs} imposes two main requirements between a concrete and abstract module. First, if $\abs M$ does not block for abstract input $\abs i$, then any concrete input $i$ associated with $\abs i$ via $\quant{\II}(i,\abs i) $ does not block; that is, the abstract component is more aggressive with rejecting invalid inputs. Second, if both systems do not block then the abstract output set is a superset of the concrete function's output set,
modulo a quantization error induced by the gridding of both inputs and outputs.
The abstract module is a conservative approximation of the concrete module because the abstraction accepts fewer inputs and exhibits more non-deterministic outputs.
Conservatism in this direction ensures that any reasoning done over the abstract models is sound and can be generalized to the concrete model. Controller synthesis tools account for blocking and non-determism in the system \cite{Tabu}. 

\begin{figure}
\centering
\input{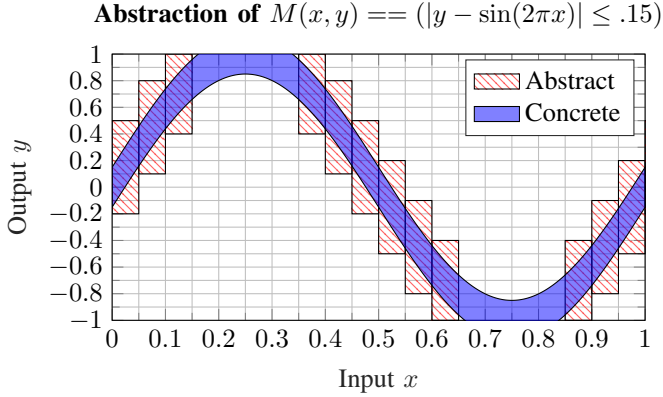}
\caption{\small The module $M(x,y) == (|y - \sin(2\pi x)| \leq .15)$ and a valid abstraction. The grid lines depict the quantization relation over a bounded domain $[0,1] \times [-1,1]$. The dark blue area represents the set of valid points for the concrete module $M(x,y)$ and the patterned red boxes represent a valid approximate abstraction $\abs M(\abs x, \abs y)$. The abstract input boxes that overlap $[.15, .35]$ or $[.65, .85]$ on the $\XX$ axis are the abstract blocking inputs $\neg \NB{\abs M}(\abs x)$. }
\end{figure} 

A feedback refinement relation (FRR) is a specialized instance of \Cref{def:relabs} for control systems, where the concrete and abstract system have identical control inputs $u$ and $\quant{u}(u,\abs u) \lequiv \top$. \Cref{eqn:FRR} is identical to the definition introduced in \cite{gunther2} but  written  as a condition on predicates. 
\begin{definition}[Feedback Refinement Relation]
Let control system modules 
%$\sys(x \cup u, x')$ and $\sys(\abs x \cup u, \abs x')$
$(\XX \cup \UU, \XX', \sys)$ and $(\abs\XX \cup \UU , \abs\XX', \abs\sys)$ 
share a common control variable $\UU$. A strict relation $\quant{\XX}$ is a feedback refinement relation from $\sys$ to $\abs \sys$ if 
\begin{align}
\quant{\XX}(x,\abs x) \AND \NB{\abs\sys}(\abs x, u) \implies \NB{\sys}(x,u)
\end{align}
is true for all assignments to $x,\abs x, u$ and
\begin{align}
\left(
	\begin{array}{c}
	\quant{\XX}(x,\abs x) \AND \NB{\abs \sys}(\abs x, u) \AND \quant{\XX}(x', \abs x') \\
	\AND \sys(x,u , x') \\
	\vimplies \\
	\abs\sys (\abs x, u, \abs x')
	\end{array}
\right)
\end{align} 
is true for all assignments to $x,\abs x, u, x',\abs x'$. Let $\abs \sys \modabs_{\quant{\XX}} \sys$ denote that $\quant{\XX}$ is a feedback refinement relation from $\sys$ to $\abs \sys$.
\label{eqn:FRR}
\end{definition} 

The most important property of feedback refinement relations is that a controller designed for an abstract system $\abs\sys$ can be refined into a controller for the concrete system. 
\begin{theorem}[Informal Statement of Theorem VI.3 from \cite{gunther2}]
If there exists a controller that enforces a behavior for abstract system $\abs \sys$ and $\abs \sys \modabs_{\quant{\XX}} \sys$ then there exists a controller for the concrete system such that the closed loop system satisfies that same behavior, modulo an approximation error with respect to state quantization $\quant{\XX}$.
\end{theorem}

The next section describes how finite abstractions are constructed and represented. It resembles existing abstraction methods for continuous state systems and is included for completeness. Readers who are familiar with symbolic control synthesis packages \texttt{PESSOA}\cite{Mazo2010} or \texttt{SCOTS}\cite{Rungger2016}, and binary decision diagrams may skip directly to \Cref{s:modcomposition}, which contains our main technical result on module composition. 

\section{Constructing Finite Module Abstractions} \label{s:absconstruction}
%Control systems in practice are typically represented in the form of an executable function or a Simulink diagram. 
This section provides background about the data structure used to store predicates over finite domains and an algorithm to construct an abstract predicate via overapproximations of forward reachable sets.

\subsection{Storage and Manipulation of Predicates}
The efficiency of constructing, storing, and manipulating abstractions depends on their underlying data structure. 
Hash tables and sparse matrices can be used to store an abstract module's input-output pairs, but require exponential memory with respect to the module input dimensionality.

The storage and manipulation problems are mitigated by representing the system and interconnection abstractions with ordered binary decision diagrams (BDDs) \cite{Bryant1986BDD}, a data structure that compactly represents predicates by detecting symmetries and redundant structure. 
%\todo{These functions represent system dynamics by returning true whenever a transition $(\hat x, \hat u, \hat w, \hat x_d)$ is valid for system $\hat \Sigma$.
%For interconnections $\hat{\mathcal{I}}= (\hat U, \hat Y, \hat M)$, the function returns true when $(\hat u, \hat y)$ satisfies the relation $\hat y \in \hat M (\hat u)$. }
BDDs are an implicit representation and often exhibit a smaller memory footprint compared to explicit representations that store every transition in memory.
The CUDD toolbox \cite{Somenzi2015} provides functionality for common operations such as taking conjunctions, disjunctions, negations, variable renaming, equality checking, and existential/universal quantification over a set of variables. These operations are performed directly on BDDs and have been empirically observed to be more memory and time efficient than analogous operations implemented for lookup tables \cite{novelbddencoding}.

\subsection{Abstraction via Module Input Space Traversal}
If quantization relations $\quant{\II}^{-1}$ and $\quant{\OO}$ are viewed as set-valued maps $\quant{\II}^{-1}: \abs \II \maps \pow{\II}$ and $\quant{\OO}: \OO \maps 2^{\abs \OO}$, then the tightest overapproximation of $F: \II \maps \OO$ can be obtained by the composition of functions\footnote{While $\quant{\II}^{-1}$ outputs subsets of $\II$ and $F$ accepts only \textit{elements} of $\II$ as inputs, we simply view the function $F$ in this case as computing the image of the set outputted by $\quant{\II}^{-1}$. Similarly, $\quant{\OO}$ computes the image of the set outputted by $F$.} $\quant{\OO}(F(\quant{\II}^{-1}(\abs i)))$. 
This composition can be viewed as a quantized version of a one step reachability problem where the image of the set of points contained in the partition $\quant{\II}^{-1}(\abs i)$ under the map $F$ is computed. 

Exact characterizations of reachable sets for arbitrary systems do not exist, but there are several practical methods to compute overapproximations  $\OA: \abs \II \maps \pow{\abs \OO}$ that satisfy the set containment condition $\quant{\OO}(F(\quant{\II}^{-1}(\abs i))) \subseteq \OA(\abs i)$. 
A few overapproximation procedures are available when a dense subset of Euclidean space is partitioned into hyper-rectangular regions. Consider a generic module $F: \mathbb{R}^m \maps \mathbb{R}^n$ and a hyper-rectangle $[a,b]$ defined by the two corners $a, b \in \mathbb{R}^m$ where $a_i \leq b_i$ for all $i = 1,\ldots,m$. Component-wise Lipschitz bounds and monotonicity are two common methods to overapproximate $F([a,b])$.
\begin{example}[{Component-wise  Lipschitz Bounds}] Let $c := \frac{1}{2}(a+b)$ and $r := \frac{1}{2} (b-a)$. Let $L$ be a matrix with non-negative entries satisfying $|F(y) - F(x)| \leq L |y-x|$ for all $x,y$ where the absolute values and inequality are component-wise. The hyper-rectangle
$[F(c) - Lr, F(c) + Lr]$ is a superset of $F([a,b])$. 
\end{example}
\begin{example}[{Monotone Functions}]
If $F$ is a monotone function, then $[F(a), F(b)]$ is a superset of $F([a,b])$. 
\end{example}
%\begin{figure}
%\caption{\todo{Figure with different overapproximation schemes.}}
%\end{figure}

\Cref{alg:func_abs} is a procedure that only uses evaluations of $OA$ to construct a finite abstraction.

\begin{algorithm}
    \caption{Module Abstraction Pseudo Code}\label{alg:func_abs}
\begin{algorithmic}[1] % The number tells where the line numbering should start
	\Function{AbstractModule}{Overapprox. $OA: \abs \II \to \pow{\abs \OO}$}
		\State $\hat F := \false$ \Comment{No transitions initially} \label{algln: initfalse} 
				\ForAll{$[a] \in \quant{\II}^{-1}(\dom{\abs \II)}$} \Comment{Input Grid Traversal} \label{algln:lowdimgrid}
				\State $T' := \false$
				\If{$OA([a])) \not\subseteq \quant{\OO}^{-1}(\dom{ \abs \OO})$} \label{algln:outofgrid}
				\State continue \Comment{Imposes blocking input}
				\EndIf 
				\ForAll{$[b] \in \quant{\OO}(OA([a]))$} \label{algln:outputfor}
					\State $T' := T' \OR (\abs o == [b])$ \Comment{Construct output set}
				\EndFor
		    	\State  $\abs{F} := \abs F \OR \left( (\abs i == [a]) \AND T' \right)$ \Comment{Adds transitions} \label{algln:append} 
		\State \Return $\abs F$ 
		\EndFor 
	\EndFunction
\end{algorithmic} 
\end{algorithm}
Note that $[a]$ and $[b]$ represent specific values of the abstract input and output sets; this value changes with each iteration of the loops in lines \ref{algln:lowdimgrid} and \ref{algln:outputfor}. 
Line \ref{algln:outofgrid}  checks for out of bounds errors, which are common when the concrete domain is Euclidean space and the abstract domain corresponds to a finite cover with bounded subsets. 
For instance, an unstable system may exit the bounded region covered by the abstract outputs for inputs that do not stabilize the system.
The continue command on the next line causes the abstract system to block for that input by preventing any transition from being added. 
Blocking inputs are accounted for within the synthesis procedure \cite{Tabu}.
If the system does not block, then the loop on line \ref{algln:outputfor} constructs the output set and line \ref{algln:append} adds the input-output pairs to the predicate. 

Line \ref{algln:lowdimgrid} of \Cref{alg:func_abs} is a nested loop with depth equal to the dimension of the input grid and is the source of the exponential runtime. Breaking apart a larger system into smaller modules and abstracting those limits the dimension of the largest module's input grid. 
Existing tools \texttt{PESSOA} \cite{Mazo2010} and \texttt{SCOTS}\cite{Rungger2016}%, and \texttt{CoSyMa}\cite{mouelhi2013cosyma}
construct symbolic representations by applying \Cref{alg:func_abs} directly on a monolithic control system.

%Recent work on multi-scale grids \cite{Hsu:2018:MAC:3178126.3178143} and partition refinement \cite{Nilsson2017} have sought to speed up the control synthesis procedure, but still adopt a monolithic view of control systems. 
%\begin{theorem}
%\Cref{alg:func_abs} applied to system updates and latent variable constraints yields a discrete counterpart that satisfies the functional abstraction relation.
%\end{theorem}
%\begin{proof}
%\todo{Proof here. Looks at all inputs to the abstract function. By definition of $\OA$ $T'$ over estimates all possible transitions that are possible. } 
%\end{proof}

%\begin{remark}
%\Cref{alg:func_abs} was stated for the scalar case, but can generalize easily. 
%\end{remark} 

\section{Module Composition} \label{s:modcomposition}
%The previous section showed that abstracting modules quickly becomes intractable as the number of module inputs increases. Instead of directly abstracting high dimensional modules, we can construct them by composing smaller ones. 
%Consider a collection of modules. 
We model connections between modules implicitly via shared variables, which represent the wires between modules and the module ports they connect. Module behaviors are coupled because variables can only assume one value. Connections between modules can be made simply by renaming variables. 
We adopt a module composition operation from \cite{tripakis2011theory} that can be applied to both concrete and abstract modules.
\newcommand\IO[1]{io_{#1}} 
\begin{definition}[Module Composition] 
Let $(\II_1, \OO_1, M_1)$ and $(\II_2, \OO_2, M_2)$ be modules with disjoint output variables $\OO_1 \cap \OO_2 = \emptyset$. Without loss of generality, let $\IO{12} := \OO_1 \cap \II_2$ and \begin{align}
\II_1 \cap \OO_2 \sequiv \emptyset \label{eq:nofeedback}
\end{align}
signifying that outputs of module $M_2$'s may not be fed back into inputs of $M_1$.
Define new composite variables
\begin{align}
\II_{12} &:= (\II_1 \cup \II_2) \setminus \IO{12} \\
\OO_{12} &:= \OO_1 \cup \OO_2 %\cup \IO{12}
\end{align}
and the composed module $(\II_{12}, \OO_{12}, M_{12})$ with predicate 
\begin{align}
M_{12} &:= M_1(i_1,o_1) \AND M_2(i_2,o_2) \label{eq:compositionpredicate} \\
& \qquad\qquad \AND \uquant{\OO_{12}}(M_1(i_1,o_1) \implies \NB{M_2}(i_2)) \nonumber
\end{align} 
The module subscripts may be swapped if instead the outputs of $M_2$ are fed into $M_1$. 
\label{def:modcomposition}
\end{definition}

We say that $M_{12}$ is a parallel composition of $M_1$ and $M_2$ if $\IO{12} \sequiv \emptyset$ holds in addition to \Cref{eq:nofeedback}. \Cref{eq:compositionpredicate} under parallel composition reduces down to $M_1 \AND M_2$ (Lemma 6.4 in \cite{tripakis2011theory}) and the composition operation is both commutative and associative. If $\IO{12} \not\sequiv \emptyset$, the modules are composed in series and the composition operation is only associative.

\begin{figure}
\centering
\includegraphics[width = .7\columnwidth]{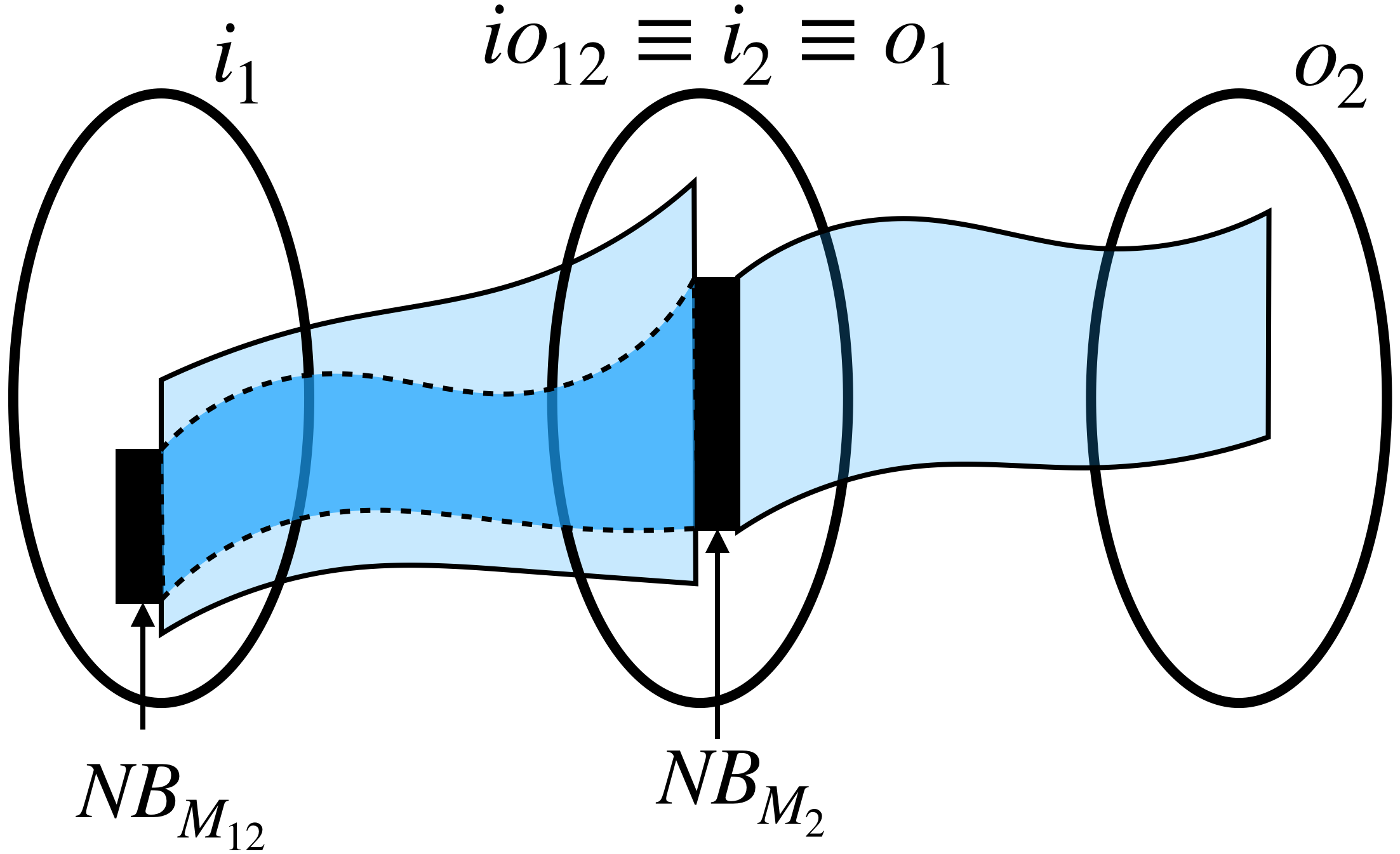}
\caption{\small Visualization of the propagation of nonblocking inputs under series composition from \Cref{def:modcomposition}. Module $M_2$'s nonblocking set $\NB{M_2}$ is a strict subset of the possible outputs of module $M_1$. Some nonblocking inputs to $M_1$ may lead $M_2$ to block instead. The additional constraint $\uquant{\IO{12}}(M_1 \implies \NB{M_2})$ causes $\NB{M_{12}}$ to prune inputs to $M_1$ that could induce $M_2$ to block under an adversarial choice of $o_1$, even though those inputs would not cause $M_1$ to block. \label{fig:NBpropagation}}
\end{figure}

The last term in \Cref{eq:compositionpredicate} is a predicate over the expanded input set $\II_{12}$ and deals with blocking behaviors under series composition. It disallows inputs whenever there exists an output of $M_1$ that causes $M_2$ to block. 
The new module's nonblocking inputs $\NB{M_{12}}$ are:
\begin{align}
\NB{M_{12}} := \equant{\OO_{12}} (M_1 \AND M_2) \AND \uquant{\OO_{12}}(M_1 \implies \NB{M_2}). \label{eq:composedNB}
\end{align}
\Cref{fig:NBpropagation} explains the role of the right-most term for a series composition of generic modules, but we also include a concrete example. 
\begin{example}
Consider a module $(x,y, M_1)$ with predicate $M_1(x,y) := (|y - x| \leq 2)$, which feeds into a module  $(y,z,M_2)$ with predicate $M_2(y,z) := (z == \sqrt{y})$. $M_2$'s nonblocking inputs $\NB{M_2}(y) $ are  $(y \geq 0)$. Substituting into the term $\equant{\OO_{12}} (M_1 \AND M_2)$ from \Cref{eq:composedNB} yields
\begin{align*}
&\equant{y}\equant{z} (|y - x| \leq 2 \AND z == \sqrt{y}) \lequiv (x \geq -2)
\end{align*}
because for any $x \geq -2$ the assignments $y := x+2$ and $z := \sqrt{x+2}$ satisfy the expression. However the serial composition is not robust to an adversarial assignment to $y$, e.g. $x = -1, y = -1$. Substituting into the term $\uquant{\OO_{12}}(M_1 \implies \NB{M_2})$ from (\ref{eq:composedNB}) yields a tighter constraint on inputs. 
\begin{align*}
& \uquant{y}\uquant{z}(|y - x| \leq 2 \implies y \geq 0)\\
& \quad \lequiv \uquant{y}( |y - x| > 2 \OR y \geq 0)\\
& \qquad \lequiv (x \geq 2). 
\end{align*}
Any input $x < 2$ is disallowed because there exists a strictly negative $y$ that satisfies $|y-x| \leq 2$.
\end{example}

\begin{algorithm}
    \caption{Compose a Collection of Modules} \label{alg:composecollection}
\begin{algorithmic}[1] % The number tells where the line numbering should start
\Function{\textsf{\comp}}{Modules $M_1, \ldots, M_N$}
\Require No algebraic loop in $G$ 
\State Construct dependency graph $G$ 
\State $s_1, \ldots, s_N$ := TopologicalSort($G$) \Comment{System Indices}
\State $\sys := (\emptyset, \emptyset,\true )$ \Comment{Initialize Empty Module}
%\State $j := 1$
%\While{$j \leq N$}
\For{($j := 1 ; j \leq N; j := j+1$)}
\State $\sys := \comp(M_{s_j}, \sys)$ \Comment{From \Cref{def:modcomposition}}
%\State $j := j + 1$
\EndFor
\State \Return $\sys$
\EndFunction
\end{algorithmic}
\end{algorithm}

\Cref{alg:composecollection} is used to compose a collection of modules through systematic application of the binary module composition operator from \Cref{def:modcomposition}.
% It is aware of restrictions on the interconnection topology via \Cref{eq:nofeedback}. 
It first constructs a directed dependency graph where all vertices are modules and an edge exists from $M_1$ to $M_2$ if $\OO_1 \cap \II_2 \neq \emptyset$ (i.e., outputs of $M_1$ feed into $M_2$). We assume that the constructed graph is directed and acyclic; this requirement ensures that there are no algebraic loops. The graph can be topologically sorted into a sequence of indices $s_1, \ldots, s_N \in \{1,\ldots, N\}$ where a directed edge from $M_{s_i}$ to $M_{s_j}$ implies that $s_j < s_i$. This would cause ``downstream" module $M_2$ to come before ``upstream" $M_1$. 
The topological sort is necessary because composing modules in any arbitrary order may violate the requirement in \Cref{eq:nofeedback} by introducing circular dependencies and feedback composition. 
Although the topological sort does not necessarily yield a unique linear module ordering, associativity and commutativity of the composition operation ensures that \Cref{alg:composecollection}'s output is unique.

Composing a large collection of variables cause the number of outputs to grow.
The output hiding operator can be used to ignore superfluous module outputs.
\begin{definition}[Output Hiding] \label{def:varhiding}
Hiding $(\II, \OO \cup w, M)$'s output $w$ yields the module $(\II,\OO ,\equant{\W}M)$.
\end{definition}
Existentially quantifying out the hidden variable ensures that the input-output behavior over the unhidden variables is still consistent with the behaviors with the hidden variable.

Control modules $(\XX \cup \UU, \XX', F)$ only have next state variables $\XX'$ as outputs. 
A control system can be obtained by composing a set of components using \Cref{alg:composecollection} and hiding all latent variables that become outputs as a result of series composition.

\subsection{Main Result: Module Composition and Output Hiding Preserve Approximate Abstraction Relations}

The following theorems show that the composition and output hiding operators preserve the relation between abstract and concrete modules. Proofs are provided in the appendix. 
\begin{theorem} \label{thm:compabs}
Let $\abs M_1 \modabs_{\II_1, \OO_1} M_1$ and $\abs M_2 \modabs_{\II_2, \OO_2} M_2$ and modules $M_{12}$ and $\abs M_{12}$ be the resulting composed modules from \Cref{def:modcomposition}. Then $\abs M_{12} \modabs_{\II_{12}, \OO_{12}} M_{12}$.
\end{theorem}

\begin{theorem} \label{thm:hiding}
If $\abs M \modabs_{\II, \OO_1 \cup \OO_2} M$ for modules $(\II, \OO_1 \cup \OO_2, M)$ and $(\abs \II, \abs \OO_1 \cup \abs \OO_2, \abs M)$, then $\equant{\abs \OO_2} \abs M \modabs_{\II, \OO_1} \equant{\OO_2} M$. 
%In other words, the output variable hiding operator preserves the module approximation condition in \Cref{def:relabs}.
\end{theorem} 

Because control systems are modules and feedback refinement relations are a special case of \Cref{def:relabs}, it readily follows that control systems can be constructed from module composition and variable hiding.  
%\begin{corollary}
%\label{thm:compcontrolabs}
%\end{corollary}

%\subsection{Suggestions for Maximizing Compostionality's Benefits}

While \Cref{thm:compabs,thm:hiding} show that abstractions are preserved under composition, unnecessarily decomposing a system may destroy structural properties and introduce additional non-determinism. For example, let a concrete module $(x,y, y == 3x)$ and its inverse $(y,z, z == \frac{y}{3})$ be composed in series and the $y$ output hidden, yielding the identity module $(x, z, z == x)$. Discretizing the variable $y$ first breaks the identity relationship between the two concrete modules, so composing abstract modules induces more non-determinism than abstracting the identity module directly.

%\Cref{alg:composecollection} is one of many possible methods to compose a collection of modules. While it composes modules by adding components one at a time in Line 7, it may be beneficial to group together small modules into slightly larger ones before composing them, as pointed out in \cite{clarke1999model}. \Cref{fig:compgroupings} depicts two methods to group modules which can informally be described as ``series composition of components in parallel" and ``parallel composition of components in series".
%Characterizing an optimal grouping is highly dependent on the topology and the modules themselves. An empirical study over a range of different use cases is one potential line of future work.
%\begin{figure} 
%\centering 
%\includegraphics[width = .5\columnwidth]{figures/twocompschemes.png} 
%\caption{\small Directed graph for the six modules in \Cref{fig:simulinkex} and two methods to compose them. The top method initially groups together parallel components then composes those groups in series. The second method generates collections of modules that are interconnected in series, then composes those groups in parallel.} \label{fig:compgroupings}
%\end{figure}

%\subsection{Constructing Abstractions over Longer Time Horizons}
%
%\todo{Series composition $x[k+2] := f(f(x[k],u[k]),u[k+1])$ for taking a one-step system and composing copies of it to get a $k-$step system. }

%\input{opspreserveabs.tex}
%\input{decompcpre.tex}
\section{Related Work} \label{sec:related}

This paper generalizes the core insights from \cite{sparseabs2017} and \cite{Kim:2018:CCS:3178126.3178144}. High dimensional system abstractions are constructed via parallel system composition \cite{sparseabs2017}, but the computational gains are most dramatic when the dynamics inherently exhibited a locality property where states are independent of one another. A limited form of series composition is introduced in \cite{Kim:2018:CCS:3178126.3178144} to further decompose the system, but it is unable to handle blocking inputs because it does not include the right-most term in \Cref{eq:compositionpredicate}. Instead of feedback refinement relations, it was based on alternating simulation relations and required that each control input be associated with a unique component. 

This paper differs from the existing literature on large-scale controller synthesis, which generally aims to solve a decentralized control problem where multiple agents each have control over a local control input and need to jointly satisfy some property. Guaranteeing satisfaction is difficult due to coupled system dynamics, concurrent decision making, and controllers only having access to local information. Many existing results certifying correctness of a decentralized controller assume a decomposed specification and rely on a certificate that the actions of an individual controller do not interfere with another controller's ability to satisfy its specification. This certificate ensures soundness of the decomposed synthesis procedure and may come in the form of a stability certificate \cite{rungger2016compositional, zamani2017compositional, mallik2016compositional}, assume-guarantee constraints \cite{meyer, kim2015compositional}.
%Constructing this certificate often requires knowing the system interconnection topology, so these results are not compositional in the sense that modules cannot be substituted for alternative components or rearranged. 
Finding a suitable system decomposition often case specific, restricting the portability of the decomposed controller synthesis results. 

%In contrast, we do not attempt to produce a collection of decentralized controllers via system decomposition. By taking a constructive approach to representing system dynamics, our approach is portable and flexible. Viewing system dynamics as a complex interconnection of simpler components could also lend itself to decomposed or parallel algorithms for controller synthesis. 
%\todo{Decomposed dynamic programming with HJI level set methods.}
%\cite{Burch:1991}

%In contrast, we do not attempt to produce a collection of decentralized controllers. Our problem is instead motivated by the inability of existing abstraction-based controller synthesis tools to scale to high dimensional systems.

%Approximate alternating simulation relations \cite{majid} are another relationship that is used for controller synthesis via finite abstraction.

%The sparsity-aware abstraction procedure from \cite{sparseabs2017} included \Cref{alg:func_abs} as a subroutine to construct an abstraction for each state update module and then composed modules in parallel. Serial composition of abstract modules was permitted in \cite{Kim:2018:CCS:3178126.3178144} but there was an implicit assumption that blocking did not occur. 
%The combination of \Cref{alg:composecollection} and \Cref{alg:func_abs} reduce down to the system abstraction procedures in \cite{sparseabs2017} and \cite{Kim:2018:CCS:3178126.3178144} if the aforementioned assumptions are satisfied. 
\section{Example}

%\subsection{Identity Dynamics}
%
%The identity dynamics are extremely simple and trivially decomposable into a set of parallel operations. This example mainly exists to demonstrate how existing controller synthesis tools run into easily avoidable bottlenecks as the state dimension grows.
%
% 

%\subsection{Composition of Bistable Systems with Broadcasted Average}
\newcommand{\sat}[3]{\text{GLOG}_{[#1,#2]} (#3)}%
We consider a set of $N$ single-input scalar systems where $\dom{\XX_i}  := [0,32]$, $\dom{u_i} := \{-2,-1,1,2\}$ for each $i = 1,\ldots, N$. 
The dynamics for the $i$-th update are given by
\begin{align}
x_i'  ==\sat{0}{32}{x_i + u_i + k(x_i - l_1)} \label{e:bimodal_sys}
\end{align}
where $k = 0.2$ and $\sat{0}{31}{\cdot}$ is a generalized logistic function with output values within the range $[0,32]$
\begin{align}
\sat{0}{32}{x} == \frac{32}{1+e^{-.2(x-\frac{b+a}{2})}}. \label{e:glog}
\end{align}
%The interconnected system needs to satisfy a stability objective where it must reach and remain within the consensus region:
%\begin{align}
%\phi = \bigvee_{\theta=0,1,\ldots, 31} \left(\bigwedge_{i=1}^N \left( \Vert x_i - \theta \Vert \leq 3 \right) \right), \label{e:invariant}
%\end{align}
%signifying that all agents have states $x_i$ that lie within a neighborhood of a common value $\theta$. 
%After the consensus region is reached, the value of $\theta$ may still change over time as the interconnected system executes. 
%If $\theta$ were a known and fixed constant then the objective would be trivially decomposable into $N$ pieces and a controller could be synthesized for each subsystem independently. 
%Because $\theta$ may vary and is not fixed beforehand, it is non-trivial to synthesize a controller for each subsystem \todo{$\Sigma_i$} individually if the controllers only have access to the local state $x_i$. 
%The reach and remain objective is succinctly expressed in temporal logic as $\Diamond \Box \phi$.
%
%Two system properties impede satisfaction of $\Diamond \Box \phi$. First, for fixed values $u_i =0$ and $w=0$, system (\ref{e:glog}) exhibits an unstable fixed point around $x_i^* = 16.0$ and two stable fixed points around \todo{$x_i^* = 1.92$ and $x_i^* = 29.07$}. 
%This bimodal property is problematic due to the divergence that occurs when different systems have initial states above and below $16.0$.
%The second difficulty in enforcing $\Diamond \Box \phi$ is due to the interaction amongst systems. 
The latent variable $l_1$ encodes the average state
\begin{align}
l_1 == \frac{1}{N} \sum_{i = 1}^N x_i . \label{e:diff_from_avg}
\end{align} 
Using \Cref{alg:func_abs} to directly construct the module (\ref{e:diff_from_avg}) requires iterating over a joint state space $\prod_{i=1}^N \dom{x_i}$. When $N =6$ this is computationally prohibitve, so we introduce two latent variables $l_2, l_3$ representing intermediate averages
\begin{align*}
l_2 == \frac{1}{3}\left(x_1 + x_2 + x_3 \right) \quad \text{ and } \quad l_3 == \frac{1}{3}\left(x_4 + x_5 + x_6 \right).
\end{align*}
The above equations and (\ref{e:combine_partial_avg}) below are equivalent to (\ref{e:diff_from_avg}).
\begin{align}
l_1 == \frac{1}{2}\left(l_2  + l_3 \right). \label{e:combine_partial_avg}
\end{align}
By construction, all latent variables lie within the space $[0,32] \sequiv \dom{l_1} \sequiv \dom{l_2} \sequiv \dom{l_3} $.
We discretize each continuous space $\XX_i$ into 32 bins and the input space is already discrete. 
%(0.0766572 + 0.0653037 +  0.0673297 + 0.0761998 + 0.0717792 + 0.0631475) + 1.11073 = 1.53 
Constructing abstractions for all nine modules takes a total of $1.53$ seconds. There are six modules $(\XX \cup \UU_i, \XX_i', F_i)$ that constrain each output $\XX_i'$. These modules are constructed  over a total of $20.35$ seconds via series composition and hiding the abstract latent variables $\abs l_1, \abs l_2, \abs l_3$. These six modules are then composed in parallel over $343.02s$ to obtain the final monolithic system which has $8.80 \times 10^{14}$ transitions. Abstracting the control system monolithically requires a traversal over a grid of $4.4 \times 10^{12} \approx (32 \times 4)^6$ state-control pairs and does not terminate after running 8 hours. All benchmarks are run on a standard laptop with $8$GB of RAM and a 2.4 GHz quad-core processor with a modified version of \texttt{SCOTS}. 
Other benchmarks with fewer discrete states and transitions have required tens of hours \cite{nilsson2016correct} and hundreds of gigabytes of RAM \cite{weber2017optimized} to compute and store abstractions. Although these benchmarks had different hardware and software implementations, they highlight how a monolithic approach to abstraction quickly grows intractable.

\appendices

\section{Proofs of \Cref{thm:compabs,thm:hiding}}

\begin{proof}[Proof of \Cref{thm:compabs}]
{
\newcommand{\JJ}{j}%{\mathbb{J}}
\newcommand{\KK}{k}%{\mathbb{K}}
\begin{figure}
\centering 
\includegraphics[width = .8\columnwidth]{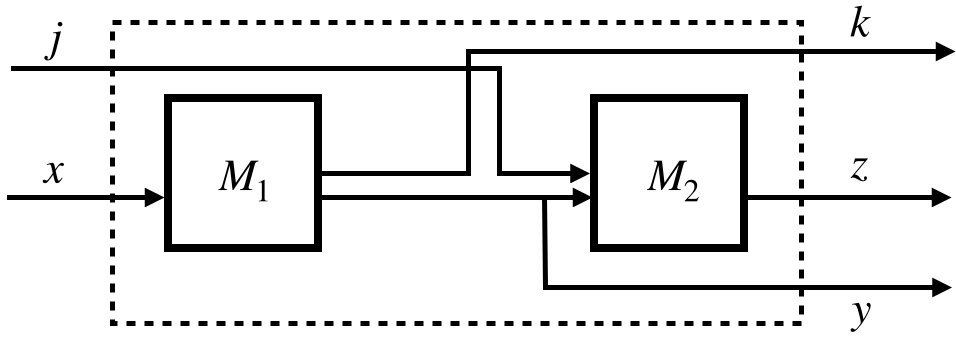}
\caption{\small Generic composition of concrete modules used in \Cref{thm:compabs}'s proof.} \label{fig:comp}
\end{figure}
Consider two modules $(\XX, \KK \cup \YY, M_1)$, $(\JJ \cup \YY, \ZZ, M_2)$. \Cref{fig:comp} depicts a general interconnection where $\IO{12} \equiv \YY$ is a subset of the outputs of $M_1$ and inputs of $M_2$. It follows that $M_{12} = (\XX \cup \JJ, \YY \cup \KK \cup \ZZ, M_{12})$ where the input-output behavior is constrained by the predicate \footnote{The quantifier over $\KK$ can be moved in because $\NB{M_{12}}$ does not depend on $\KK$. The universal quantifier over $\ZZ$ can be removed altogether because $M_1$ and $\NB{M_{2}}$ do not depend on $\ZZ$. }
\begin{align}
M_{12} :=\;& M_1 \AND M_2 \AND \uquant{\YY}\uquant{\KK}\uquant{\ZZ}(M_1 \implies \NB{M_2})\\
:=\;& M_1 \AND M_2 \AND \uquant{\YY}(\uquant{\KK}M_1 \implies \NB{M_2}).
\end{align} 
Predicate $\NB{M_{12}}(\XX, \JJ)$ encodes the set of nonblocking inputs:
\begin{align*}
\NB{M_{12}} &:= \equant {\YY}\equant{\KK}\equant{\ZZ}(M_1 \AND M_2 \AND \uquant{\YY}(\uquant{\KK}M_1 \implies \NB{M_2}))\\
%&:=\equant{\YY}(\equant{\KK}M_1 \AND \equant{\ZZ}{M_2}) \AND \uquant{\YY}(\uquant{\KK} M_1 \implies \NB{M_2})  \\
&:=\equant{\YY}(\equant{\KK}M_1 \AND \NB{M_2}) \AND \uquant{\YY}(\uquant{\KK}M_1 \implies \NB{M_2}). 
\end{align*}

From \Cref{thm:compabs}'s assumptions it is known that $\abs M_1 \modabs_{\XX,\KK\cup\YY} M_1$ and $\abs M_2 \modabs_{\JJ \cup \YY, \ZZ} M_2$. These conditions written explicitly are the overapproximation condition on $M_1$
\begin{align}
\begin{array}{c}
\quant{\XX \cup \YY \cup \KK}(x,\abs x, y, \abs y, k, \abs k) %\AND \quant{\YY}(y,\abs y) \AND \quant{\KK}(k, \abs k) 
 \AND \NB{\abs M_1}(\abs x) \AND M_1(x,k,y)\\
\vimplies \\
\abs M_1(\abs x, \abs k, \abs y)
\end{array} \label{eq:oam1}
\end{align}
the overapproximation condition on $M_2$
\begin{align}
\begin{array}{c}
\quant{\JJ \cup \YY \cup \ZZ}(j,\abs j, y, \abs y, z, \abs z)% \AND \quant{\YY}(y,\abs y) \AND \quant{\ZZ}(z, \abs z) \\
\AND \NB{\abs M_2}(\abs j, \abs y) \AND M_2(j,y,z)\\
\vimplies\\
\abs M_2(\abs j, \abs y, \abs z)
\end{array} \label{eq:oam2} 
\end{align}
and the nonblocking conditions
\begin{align}
\quant{\XX}(x,\abs x) \AND \NB{\abs M_1}(\abs x) &\implies \NB{M_1} (x) \label{eq:nonbm1}\\
\quant{\JJ}(j,\abs j) \AND \quant{\YY}(y,\abs y) \AND \NB{\abs M_2}(\abs j, \abs y) &\implies \NB{M_2} (j,y). \label{eq:nonbm2}
\end{align}

We first prove the overapproximation condition from \Cref{def:relabs} for $M_{12}$.
\begin{align}
\begin{array}{c}
\quant{\XX \cup \JJ \cup \YY \cup \KK \cup \ZZ}(x,\abs x, j, \abs j, y, \abs y, k, \abs k, z, \abs z) \AND M_{12}(x,j,y,k,z)\\ % \AND \quant{\JJ}(j,\abs j) \AND \quant{\YY}(y,\abs y) \AND \quant{\KK}(k,\abs k) \AND \quant{\ZZ}(z,\abs z)\\
\AND \NB{\abs M_{12}}(\abs x, \abs j) \\
\vimplies \\
\abs M_1(\abs x, \abs k, \abs y) \AND \abs M_2 (\abs j, \abs y, \abs z) \AND \uquant{\abs \YY} (\uquant{\abs \KK} \abs M_1 \implies \NB{\abs M_2})
\end{array} \label{eq:compoverapprox}
\end{align} 
Suppose that all concrete and abstract variables are assigned such that the upper half of \pref{eq:compoverapprox} is true. The term $\uquant{\abs \YY} (\uquant{\abs \KK} \abs M_1 \implies \NB{\abs M_2})$ follows directly from $\NB{\abs M_{12}}(\abs x, \abs j)$.  Satisfaction of $\NB{\abs M_{12}}(\abs x, \abs j)$ implies satisfaction of $\NB{\abs M_1}(\abs x)$, which via \Cref{eq:oam1} implies satisfiaction of $\abs M_1(\abs x, \abs k, \abs y)$. Satifaction of $ \NB{\abs M_2}(\abs j, \abs y)$ can now be established because $\abs M_1(\abs x, \abs k, \abs y)$ is true and $\abs x$ and $\abs j$ are assigned to value where $\uquant{\abs\YY}(\uquant{\abs\KK} \abs M_1 \implies \NB{\abs M_{2}})$ and the specific value of $\abs y$ is irrelevant. The upper half of \pref{eq:oam2} is now satisfied, implying that $\abs M_2(\abs j, \abs y, \abs z)$ is true. We have proven the bottom half of \pref{eq:compoverapprox}. 

We next prove the nonblocking condition for $M_{12}$
\begin{align}
\begin{array}{c} 
\quant{\XX}(x,\abs x) \AND \quant{\JJ}(j,\abs j) \\
\AND \equant{\abs \YY}(\equant{\abs \KK} \abs M_1 \AND \NB{\abs M_2})  \AND \uquant{\abs\YY}( \uquant{\abs \KK} \abs M_1 \implies \NB{\abs M_2}) \\
\vimplies \\
\equant{\YY} (\equant{\KK}M_1 \AND \NB{ M_2} ) \AND \uquant{\YY}(\uquant{\KK} M_1 \implies \NB{M_2})
\end{array}\label{eq:compnonblock} 
\end{align} 
Suppose that \pref{eq:compnonblock} does not hold, which means that the following formula is satisfiable. 
\begin{align}
&\quant{\XX}(x,\abs x) \AND \quant{\JJ}(j,\abs j) \label{eq:negcompnonblock} \\ 
&\AND \equant{\abs \YY}(\equant{\abs \KK} \abs M_1 \AND \NB{\abs M_2})  \AND \uquant{\abs\YY}( \uquant{\abs \KK} \abs M_1 \implies \NB{\abs M_2}) \nonumber \\ 
&\AND \big( \uquant{\YY} (\uquant{\KK} \neg M_1 \OR \neg \NB{ M_2} ) \OR \neg \uquant{\YY}(\uquant{\KK} M_1 \implies \NB{M_2}) \big) \nonumber 
\end{align}
We show that any satisfying assignment leads to a contradiction. Let $x,\abs x, j, \abs j$ be an assignment such that \pref{eq:negcompnonblock} is true. The premises of \pref{eq:nonbm1} are implied by the clauses $\quant{\XX}(x,\abs x)$ and $\equant{\abs \YY}(\equant{\abs \KK}\abs M_1 \AND \NB{\abs M_2})$ and thus there exist assignments $x,y,k$ such that $M_1(x,k,y)$ is satisfied. 

We prove that $\uquant{\YY}(\uquant{\KK} M_1 \implies \NB{M_2})$ must be false. If it is true, then $\uquant{\YY} (\uquant{\KK} \neg M_1 \OR \neg \NB{ M_2} )$ must be hold for \pref{eq:negcompnonblock} to be satisfied. For the assignments $x,j,k,y$ such that $M_1(x,k,y)$ is true, these two statements cannot both hold since the former implies $\NB{M_2}$ while the latter implies $\neg \NB{M_2}$. 

Our problem is now simplified to proving that
\begin{align}
&\quant{\XX}(x,\abs x) \AND \quant{\JJ}(j,\abs j) \label{eq:negcompnonblocksimple} \\
&\AND \equant{\abs \YY}(\equant{\abs \KK} \abs M_1 \AND \NB{\abs M_2})  \AND \uquant{\abs\YY}( \uquant{\abs \KK} \abs M_1 \implies \NB{\abs M_2}) \nonumber \\ 
&\AND (\neg \uquant{\YY}(\uquant{\KK} M_1 \implies \NB{M_2})) \nonumber 
\end{align}
is a contradiction. The prior assignments to $x,\abs x, j, \abs j$ satisfy \Cref{eq:negcompnonblocksimple} whose last clause implies that there must exist assignments $y,k$ such that $M_1 \AND \neg \NB{M_2}$. Invoking strictness of $\quant{\KK}$ and $\quant{\YY}$, allows us to assign $\abs k, \abs y$ such that $\quant{\KK}(k, \abs k)$ and $\quant{\YY}(y, \abs y)$ are true and use \Cref{eq:oam1} such that $\abs M_1 (\abs x, \abs k, \abs y)$ is satisfied. Given all the assigned variables, we can now conclude that $\NB{\abs M_2}$ holds because of $\uquant{\abs\YY}( \uquant{\abs \KK} \abs M_1 \implies \NB{\abs M_2})$. 
However the assignment should also satisfy $\neg \NB{M_2}$, which implies through the contrapositive of \Cref{eq:nonbm2} that $\neg \quant{\JJ}(j,\abs j) \OR \neg \quant{\YY}(y,\abs y)\OR \neg \NB{\abs M_2}$. Variables $j,\abs j, y,\abs y$ were already assigned so $\quant{\JJ}(j,\abs j)$ and $\quant{\YY}(y, \abs y)$ are true, so $\NB{\abs M_2}$ must be false. Contradiction. 

%The premises of \Cref{eq:nonbm1} are implied by the clauses $\quant{\XX}(x,\abs x)$ and $\equant{\abs \YY}(\equant{\abs \KK}\abs m_1 \AND \NB{\abs m_2})$ and thus there exist assignments $x,y,k$ such that $m_1(x,k,y)$ is satisfied. Invoking strictness of $\quant{\KK}$ and $\quant{\YY}$, allows us to assign $\abs k, \abs y$ and use \Cref{eq:oam1} such that $\abs m_1 (\abs x, \abs k, \abs y)$ is satisfied. The premise $\uquant{\abs\YY}(\abs m_1 \implies \NB{\abs m_2})$ implies via \Cref{eq:nonbm2} that both $\equant{ \YY}(\equant{\KK} m_1 \AND \NB{ m_2} )$ and $\uquant{\YY}(m_1 \implies \NB{m_2})$ hold. 
}
\end{proof}
\begin{proof}[Proof of \Cref{thm:hiding}]
Substitute \Cref{def:varhiding} of a hidden variable module into the conditions from \Cref{def:relabs}. The nonblocking condition (\ref{eqn:NB}) only pertains to inputs and so remains unchanged after substitution. Substitution into the overapproximation condition (\ref{eqn:superset}) yields the constraint. 
\begin{align}
\left(\begin{array}{c} 
\quant{\II}(i,\abs i) \AND \NB{\abs M}(\abs i)  \AND \equant{\OO_2}.M(i,o_1,o_2) \AND \quant{\OO_1}(o_1,\abs o_1)\\
\vimplies \\
\equant{\abs \OO_2}.\hat{M}(\abs i, \abs o_1, \abs o_2) 
\end{array}\right) \label{eqn:hidingsuperset}
\end{align}
Consider assignments to variables $i,\abs i, o_1, \abs o_1$ such that the upper half of (\ref{eqn:hidingsuperset}) holds. Satisfaction of $\equant{\OO_2}.M(i,o_1,o_2)$ implies that there must exist an assignment to $o_2$ such that $M(i,o_1,o_2)$ holds. Strictness of $\quant{\OO_2}$ implies existence of an assignment to $\abs o_2$ such that $\quant{\OO_2}(o_2, \abs o_2)$ holds. Consider any such assignment. By hypothesis $\abs M \modabs_{\II, \OO_1 \cup \OO_2} M$, and all the variables have been assigned to imply satisfaction of $\hat{M}(\abs i, \abs o_1, \abs o_2)$. It follows that $\equant{\abs \OO_2}.\hat{M}(\abs i, \abs o_1, \abs o_2) $ is satisfied.
\end{proof}

%\section*{Acknowledgments}

\bibliographystyle{abbrv}
\bibliography{references}

% Can use something like this to put references on a page
% by themselves when using endfloat and the captionsoff option.
\ifCLASSOPTIONcaptionsoff
  \newpage
\fi

%\begin{IEEEbiography}{Eric S. Kim}
%Biography text here.
%\end{IEEEbiography}
%
%\begin{IEEEbiography}{Eric S. Kim}
%Biography text here.
%\end{IEEEbiography}
%
%\begin{IEEEbiography}{Eric S. Kim}
%Biography text here.
%\end{IEEEbiography}

% insert where needed to balance the two columns on the last page with biographies
%\newpage

% that's all folks
\end{document}